\documentclass[12pt]{iopart}

\usepackage{graphicx}
\usepackage{bm}
\usepackage{epstopdf}
\usepackage{amssymb}
\usepackage{color}
\usepackage[colorlinks=true, letterpaper=true, pdfstartview=FitV, linkcolor=blue, citecolor=blue, urlcolor=blue]{hyperref}
\usepackage{lineno}
\begin{document}

\title[]{Modulation of heat transport in two-dimensional group-III chalcogenides}

\author{Wenhui Wan, Ziwei Song, Shan Zhao, Yanfeng Ge, and Yong Liu*}

\address{State Key Laboratory of Metastable Materials Science and Technology $\&$
Key Laboratory for Microstructural Material Physics of Hebei Province,
School of Science, Yanshan University, Qinhuangdao, 066004, P.R. China}
\eads{\mailto{ycliu@ysu.edu.cn}, \mailto{yongliu@ysu.edu.cn}}
\vspace{10pt}
\begin{indented}
\item[]May 2019
\end{indented}

\begin{abstract}
  We systematically investigated the modulation of heat transport of experimentally accessible two-dimensional (2D) group-III chalcogenides by first-principles calculations.
  It was found that intrinsic thermal conductivity ($\kappa$) of chalcogenides MX (M = Ga, In; X = S, Se) were desirable for efficient heat dissipation. Meanwhile, we showed that the long-range anharmonic interactions played an important role in heat transport of the chalcogenides. The difference of $\kappa$ among the 2D group-III chalcogenides can be well described by the Slack model and can be mainly attributed to phonon group velocity. Based on that, we proposed three methods including strain engineering, size effect and making Janus structures to effectively modulate the $\kappa$ of 2D group-III chalcogenides, with different underlying mechanisms. We found that tensile strain and rough boundary scattering could continuously decrease the $\kappa$ while compressive strain could increase the $\kappa$ of 2D group-III chalcogenides. On the other side, the change of $\kappa$ by producing Janus structures is permanent and dependent on the structural details. These results provide guilds to modulate heat transport properties of 2D group-III chalcogenides for devices application.
\end{abstract}

\noindent{}
%

\section{Introduction}
Two-dimensional (2D) materials with high carrier mobility ($\mu$) have great potential applications in future
electronic, optoelectronic and thermoelectric devices~\cite{liao2010high,Phosphorene2014,zhang2015high}. These applications are inevitably related to heat management and heat rectification in devices. A high thermal conductivity ($\kappa$) is requested for fast heat dissipation of nanoscale electronic devices~\cite{zou2001}. Contrary to that, nanoscale thermoelectric devices with high conversion efficiency require a high $\mu$ and a low $\kappa$ at the same time~\cite{silinano1}. Thus, the understanding and modulation of thermal transport properties of 2D materials with high $\mu$ are of technological importance for relevant device performance.

\begin{figure}[tbp!]
\centerline{\includegraphics[width=0.6\textwidth]{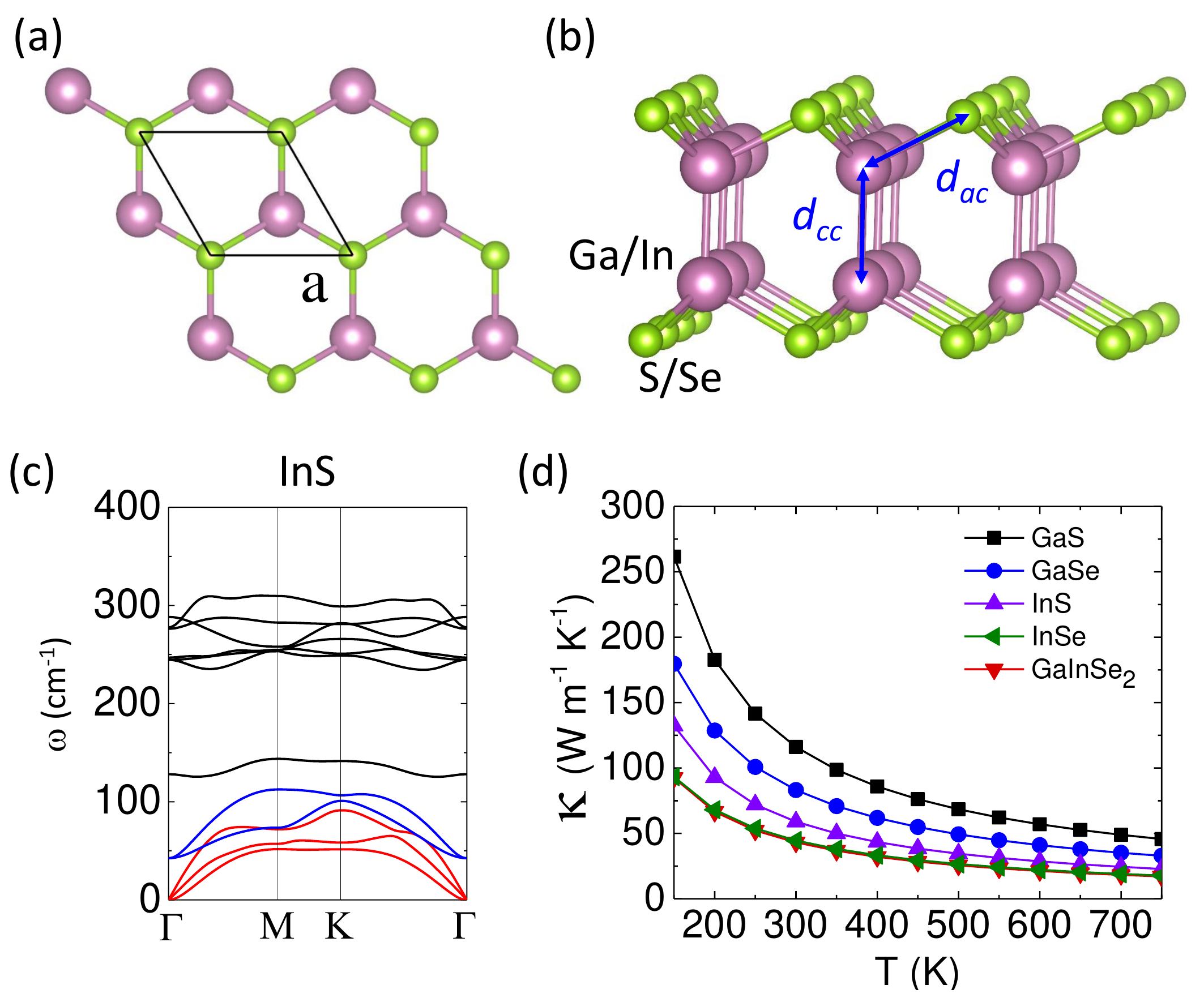}}
\caption{(a) A top view and (b) a side view of lattice crystal of 2D group-III chalcogenides with the black line showing the unit cell. The blue arrow label the cation-cation bond length $d_{cc}$ and anion-cation bonding length $d_{ac}$. (c) Phonon dispersive relation of monolayer InS. The red, blue and black lines represent the acoustic, low-frequency optical and high-frequency optical branches, respectively. (d) The temperature dependence of $\kappa$ of monolayer group-III chalcogenides and Janus monolayer GaInSe$_{2}$.}
\label{wh1}
\end{figure}

The well-known graphene~\cite{Geim1530} and 2D MoS$_{2}$~\cite{baugher2013intrinsic} suffers from the lack of a natural band gap and the low $\mu$, respectively. Phosphorene has a high $\mu$, but its $\mu$ will degrade at atmosphere condition~\cite{Wang_2016}.
Group-III chalcogenides including GaS, GaSe and InSe are hexagonal layered semiconductor and consists of quadruple sublayers (see Fig.~\ref{wh1}(b)) which are held together by van der Waals (vdW) interactions~\cite{C5CE01986A}.
Group-III chalcogenides have attracted great interest recently due to their superior properties, such as direct band gap, small effective mass, rare p-type electronic behaviors, high charge density, and so on~\cite{C6NR05976G}. Recently, 2D InSe were demonstrated to have a high electron $\mu$($>10^{3}$ cm$^{2}$/(V$\cdot$s))~\cite{bandurin2016high,InSehigh2} which is comparable with its bulk counterpart~\cite{segura1984electron}.
On the other side, the bendable photodetectors based on 2D GaS~\cite{yang2014high}, GaSe~\cite{GaSemonolayer} and InSe~\cite{tamalampudi2014high} have a ultrahigh photoresponsivity, detectivity and a broadband spectral response on flexible and transparent mica substrates, regardless of repeated bending~\cite{C5CE01986A}.
Moreover, nano-devices base on 2D InSe and GaSe have demonstrated to have good ambient stability~\cite{stable2,GaSestable}.

These advantages have intrigued many works on the electrical and optoelectronic properties of 2D group-III chalcogenides~\cite{ho2016thickness,C7NR09486H,dey2014mechanism,cao2015tunable,jin2017engineering,debbichi2015two}.
It also call for a systematical investigation to phonon transport properties and its modulation of 2D group-III chalcogenides, which is still limited~\cite{InSe,InSe2}. Besides, though 2D InS with InSe-like structure has been synthesized, its thermal properties have not been investigated~\cite{hollingsworth2000catalyzed}. In this work, we studied the heat transport of monolayer group-III chalcogenides by first-principles calculations, coupled with phonon Boltzmann transport equation (BTE). We first calculated the intrinsic thermal conductivity $\kappa$ of chalcogenides MX (M = Ga, In; X = S, Se) and found their $\kappa$ can be well described by the Slack model. Meanwhile, we showed the important role of long-range anharmonic interaction in the heat transport of group-III chalcogenides. Based on that, taking monolayer InSe as an example, we displayed three effective approaches to modulate $\kappa$ of 2D group-III chalcogenides including strain engineering, nanostructuring, and producing Janus structure.

\section{Methodology}
Based on the phonon BTE, lattice thermal conductivity $\kappa$ is calculates by~\cite{li2014shengbte}
\begin{eqnarray} \label{kappa}
\kappa_{\alpha\beta}= \frac{1}{N\Omega}\sum\limits_{
\mathbf{q},s} {C_{\mathbf{q},s}v_{\mathbf{q},s}^{\alpha}v_{\mathbf{q},s}^{\beta}\tau_{\mathbf{q},s}},
\end{eqnarray}
where $N$ and $\Omega$ is the number of $\mathbf{q}$ point and volume of the unit cell, respectively. $\alpha$ and $\beta$ are Cartesian indices. $C_{\mathbf{q},s}$, $v_{\mathbf{q},s}^{\alpha}$ and $\tau_{\mathbf{q},s}$ is the mode specific capacity, phonon group velocity and lifetime of the phonon mode with wavevector $\mathbf{q}$ and branch index $s$, respectively. The scattering mechanism to estimate phonon lifetime $\tau_{\mathbf{q},s}$ includes the anharmonic scattering ($1/ \tau_{\mathbf{q},s}^{an}$), isotopic impurities scattering ($1/\tau_{\mathbf{q},s}^{iso}$) and boundary roughness scattering $1/\tau_{\mathbf{q},s}^{b}=|v_{\mathbf{q},s}|/L$ with $L$ be the sample size. The computational details are given in the supplemental material.

\section{Results and discussion}
\subsection{The thermal conductivity of monolayer group-III chalcogenides}
Figure~\ref{wh1}(a) and~\ref{wh1} (b) display the crystal structure of monolayer group-III chalcogenides, which has the $D_{3h}$ crystal symmetry. The optimized lattice constants, bond lengths (labeled in Fig.~\ref{wh1}(b)) and thickness of monolayer are listed in Table S1 in the supplementary material. 2D group-III chalcogenides are semiconductors~\cite{C6NR05976G} and the heat is mainly carried by lattice vibrations. The phonon dispersion of monolayer InS and other materials are displayed in Fig.~\ref{wh1}(c) and Fig. S1 to display its lattice stability. The acoustic branches (see red line in Fig.~\ref{wh1}(c)) consist of the in-plane longitudinal (LA) branch, transverse (TA) branch and out-of-plane flexural branch (ZA) branch.

\begin{figure}[tbp!]
\centerline{\includegraphics[width=0.6\textwidth]{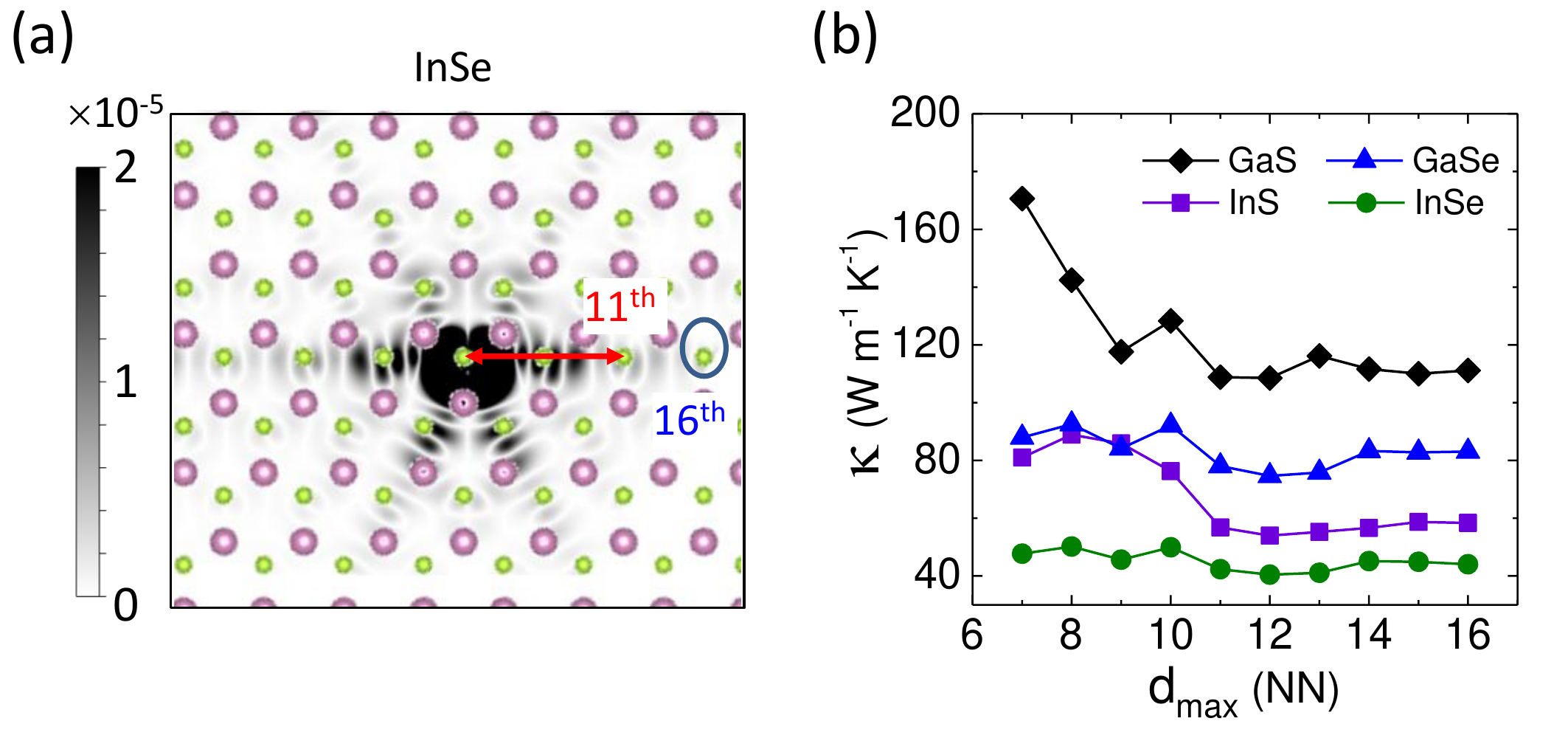}}
\caption{(a) The absolute change of charge density $|\Delta \rho|$ (in unit of $10^{-5}$ e/\AA$^{3}$) due to the small displacement (0.02 \AA) of the central Se atom along the zigzag direction. The red arrow and circle represents the distance of the $11^{th}$- and $16^{th}$-nearest neighbors with respect to the central Se atoms, respectively. (b) Lattice thermal conductivity of monolayer group-III chalcogenides at 300 K as a function of the $d_{max}$ which is in unit of nearest neighbors (NN).}
\label{wh2}
\end{figure}

We found that long-range  anharmonic interatomic force constants ($3^{rd}$ IFCs) is a common feature in monolayer group-III chalcogenides. It can be directly reflected by the response of charge density to perturbations of the atomic displacements. Taking InSe as an example, we induced a small distortion ($\sim 0.02$ \AA) to the central Se atoms along the zigzag or armchair direction. As a result, the disturbance of the absolute charge density $|\Delta \rho|$ is particularly extended along the zigzag direction, as shown in Fig.~\ref{wh2}(a).
A clear $|\Delta \rho|$ at the zone with distance from origin Se atoms less than of 8.19 {\AA} which corresponds to $11^{th}$ nearest neighbors (NN). Furthermore, the non-negligible $|\Delta \rho|$ can take place as long as 10 \AA\ ($16^{th}$ NN), as shown in Fig.~\ref{wh2}(a). Similar to our previous work~\cite{Wan_2019}, for the $3^{rd}$ IFCs of four group-III chalcogenides, we extracted its maximum component of $(\Phi^{\alpha,\beta,\gamma}_{i,j,k})_{max}$ and the maximum interatomic distance ($d_{max}$) between three atoms ($i$, $j$, $k$)~\cite{li2015orbitally}. The distribution of $(\Phi^{\alpha,\beta,\gamma}_{i,j,k})_{max}$ with respect to $d_{max}$ is shown in Fig. S2. We identified that the triplets $(\Phi^{\alpha,\beta,\gamma}_{i,j,k})_{max}$ are particularly large at the $2^{nd}$ and $11^{th}$ NN for all the group-III chalcogenides. Meanwhile, the populations of the $3^{rd}$ IFCs as the function of $d_{max}$ are shown in the inset of Fig. S2. We found that though the triplets $(\Phi^{\alpha,\beta,\gamma}_{i,j,k})_{max}$ at $14^{th}$ NN are small, it has a particularly large population. The distribution of $3^{rd}$ IFCs is consistent with the aforementioned distribution of $|\Delta \rho|$.
The convergence test of the $\kappa$ of monolayer group-III chalcogenides shows that $\kappa$ happens a clear change (see Fig.~\ref{wh2}(b)) when we include the $3^{rd}$ IFCs within the $d_{max}$ of $11^{th}$ neighbors and $14^{th}$ neighbors in the calculation of anharmonic relaxation time $\tau_{\mathbf{q},s}^{an}$. That can be understood from the aforementioned large size and large population of $3^{rd}$ IFCs at these distances. The $\kappa$ can reach convergence at $d_{max}$ of $16^{th}$ NN (see Fig.~\ref{wh2}(b)), consistent with previous charge density to perturbations of the atomic displacements.

The intrinsic $\kappa$ of monolayer group-III chalcogenides without boundary scattering are calculated at different temperatures (T), as shown in Fig.~\ref{wh1}(d). It corresponds $1/\tau_{\mathbf{q},s}=1/ \tau_{\mathbf{q},s}^{an}+1/\tau_{\mathbf{q},s}^{iso}$ according to the Matthiessen rule. The overall
of $\kappa$ is in the decreasing order of GaS, GaSe, InS and InSe.
At T = 300 K, the $\kappa$ of monolayer GaS, GaSe, InS and InSe is estimated to be 111.0, 83.10, 58.9 and 44.9 W/(m$\cdot$K), which is compatible with the previous work~\cite{InSe2}. The $\kappa$ of monolayer group-III chalcogenides are comparable with that of conventional bulk Ge (65 W/(m$\cdot$K)~\cite{Gekappa}) and GaAs (45 W/(m$\cdot$K)~\cite{Gekappa}), and larger than $\kappa$ of phosphorene which is about 20 W/(m$\cdot$K) along zigzag direaction~\cite{Luo2015,phosphorene}. Therefore, the intrinsic $\kappa$ of 2D group-III chalcogenides are desirable for efficient heat dissipation.

\Table{\label{table1}The average atomic mass $\overline{m}$ in atomic mass unit (amu), 2D effective elastic module $C_{\rm 2D}$ (J/m$^{2}$), Debye temperature $\theta_{D}$ (K) and Gr\"{u}neisen parameter $\gamma$ of monolayer group-III chalcogenides.}
\br
        &$\overline{m}$&$C_{\rm 2D}$&$\theta_{D}$&$\gamma$ \\
\mr
GaS     &50.90     &81.32 &136.5 &0.63 \\
GaSe	&74.34	   &69.46 &105.9 &0.58 \\
InS     &73.45	   &56.50 &90.1  &0.45 \\
InSe	&96.89	   &49.21 &76.3  &0.51 \\
\br
\endTable

To understand the physical factors that affect heat transport of chalcogenides, we adopted the Slack model which gives $\kappa$ as~\cite{slack1973nonmetallic},
\begin{eqnarray} \label{slack1}
\kappa = B \cdot \frac{\overline{m} \theta_{D}^{3}n^{\frac{1}{3}} \delta}{\gamma^{2}T},
\end{eqnarray}
where B is a numerical coefficient; $\overline{m}$ is the average atomic mass; $n$ is the number of atoms in the unit cell; $\delta^{3}$ gives the volume per atom; $\theta_{D}$ and $\gamma$ is the acoustic Debye temperature and Gr\"{u}neisen parameter, respectively. The simulation details are given in the supplementary material. The results are shown in Table~\ref{table1}.

A large acoustic $\theta_{D}$ reflects a large bandwidth of acoustic branches and a large phonon group velocity.
Based on a diatomic chain model which contains two types of atoms per unit cell, the mass of heavy atom and bonding stiffness have a negative and positive effect on the acoustic group velocity.
In 2D group-III chalcogenides, the bonding stiffness can be described by 2D elastic module $C_{\rm 2D}$~\cite{eleastic}.
From GaS to InSe, the mass of the heavy atom and $C_{\rm 2D}$ is in the increasing and decreasing order, respectively (see Table~\ref{table1}). Thus, the overall order of $v_{\mathbf{q},s}$ of acoustic phonons from high to low is in the sequence of GaS, GaSe, InS and InSe (see Fig. S4(a)), which agrees with the order of $\theta_{D}$ (see Table~\ref{table1}). The Gr\"{u}neisen parameter $\gamma$, however, is in the increasing sequence of InS, InSe, GaSe and GaS, which is almost opposite to order of $\theta_{D}$.

Substituting these results into Eq.~\ref{slack1}, the ratio of room-temperature $\kappa$ of monolayer GaS, GaSe, InS and InSe is $1.98:1.64:1.32:1$, consistent with that obtained by phonon BTE which is $2.47:1.85:1.31:1$. The deviation arises from the neglecting the contribution of low-frequency optical branches (labeled by blue in Fig.~\ref{wh1}(c)) to $\kappa$ in slack model whereas those phonon modes make a non-negligible contribution to $\kappa$  of chalcogenides (see Fig. S3). Thus, the $\theta_{D}^{3}$ in Eq.~\ref{slack1} ultimately overtake the $\gamma^{2}$ and other factors, dominating the overall behaviour of $\kappa$ of 2D group-III chalcogenides.

From the other side, we also performed a detailed analysis of mode contribution $\kappa_{\mathbf{q},s}$ to total $\kappa$. The $\kappa_{\mathbf{q},s}$ is given by $\kappa_{\mathbf{q},s}=C_{\mathbf{q},s}v^{2}_{\mathbf{q},s}\tau_{\mathbf{q},s}$, where $C_{\mathbf{q},s}$, $v_{\mathbf{q},s}$ and $\tau_{\mathbf{q},s}$ are specific heat per unit volume, group velocities and lifetimes of phonon mode with momentum $\mathbf{q}$ and branch index $s$, respectively.
$\tau_{\mathbf{q},s}$ is determined by both the Gr\"{u}neisen parameter $\gamma_{\mathbf{q},s}$ and phase space for phonon anharmonic scattering $(P_{3})_{\mathbf{q}, s}$~\cite{lindsay2008three}.
In the low-frequency zone, however, there is a competition between $\gamma_{\mathbf{q},s}$ and $(P_{3})_{\mathbf{q}, s}$. This makes the $\tau_{\mathbf{q},s}$ of low-frequency phonons in different chalcogenides have almost the same order of magnitude (see Fig. S4(b-d)). Meanwhile, $C_{\mathbf{q},s}$ approaches the classic value $k_{B}$ at a temperature higher than $\theta_{D}$. Thus, the difference of $\kappa$ among chalcogenides is mainly determined by phonon group velocity, which is consistent with previous analysis by Slack model.

\begin{figure}[tbp!]
\centerline{\includegraphics[width=0.6\textwidth]{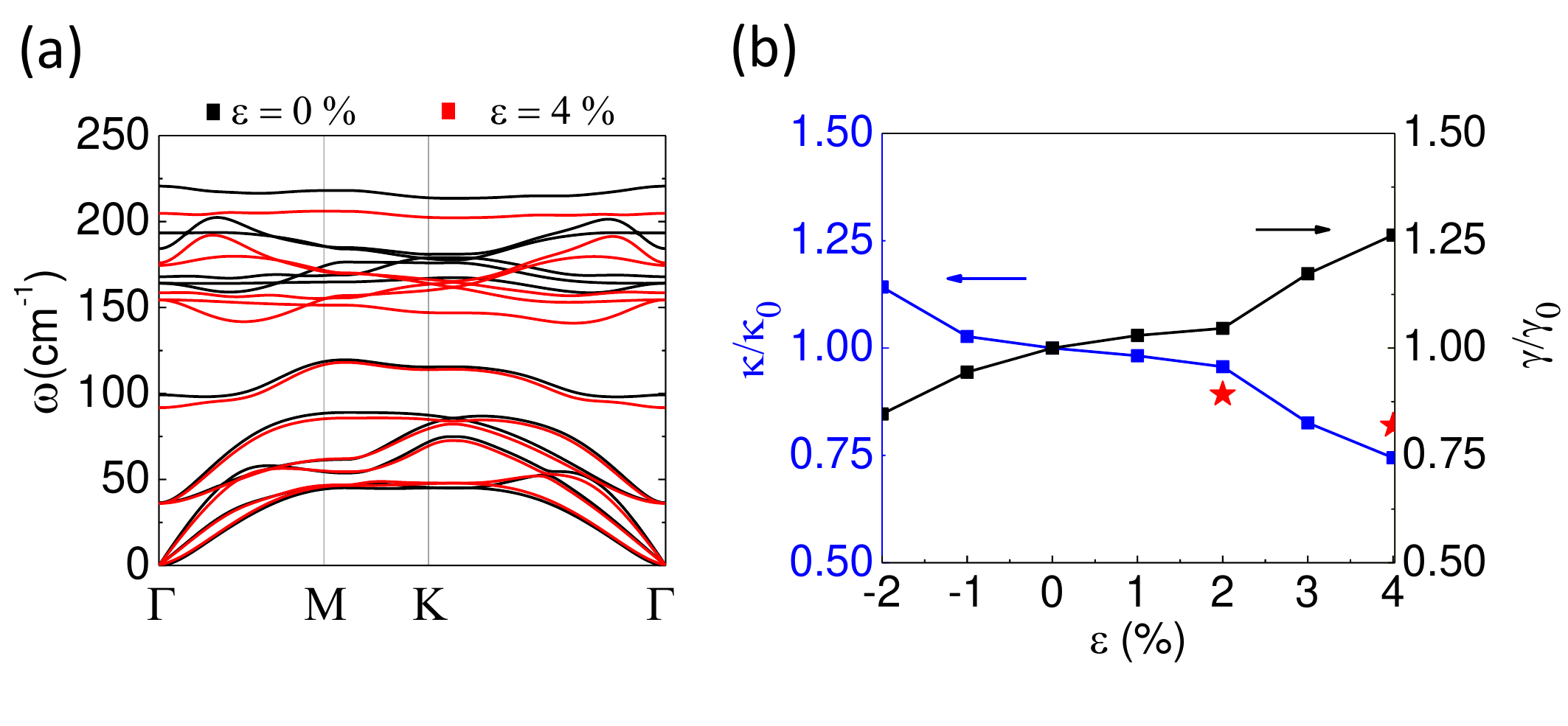}}
\caption{(a) The phonon dispersion of  monolayer InSe at tensile of ¦Å = 0\% and 4\% tensile strain. (b) Strain dependence of normalized $\kappa$ and Gr\"{u}neisen parameter $\gamma$ with respect to that of strain-free monolayer InSe according to Slack model. The ref pentagram represents the $\kappa$ gotten through the phonon BTE equation.}
\label{wh3}
\end{figure}

\subsection{The methods to modulate the lattice conductivity}

Thermal engineering of 2D material is significant to improve the transport properties and expand its application
perspective. Considering group-III chalcogenides share common features in structural symmetry, IFCs and thermal transport model, in the following, we only considered
feasible methods to modulate the $\kappa$ of 2D InSe, the conclusion will be the same with other group-III chalcogenides.

\subsubsection{Strain engineering}

Experimentally, strain engineering can be utilized to manipulate the heat transport of 2D materials~\cite{MoS2strain}. we defined the strain as $\varepsilon=(a_{i}-a_{0})/a_{0}\times 100\%$, where $a_{i}$ and $a_{0}$ is the lattice constants of monolayer InSe with and without strain, respectively. When applying tensile strain, both $d_{cc}$ and $d_{ac}$ are enlarged and the bonding are weakened. The phonon dispersion curves downshift except that the frequency of ZA phonons is slightly enhanced (see Fig.~\ref{wh3}(a)). This leads to overall smaller phonon group velocities and smaller Debye temperature $\theta_{D}$.
We calculated the parameters of Eq.~\ref{slack1} at both tensile strain and compressive strain.
Gr\"{u}neisen parameter $\gamma$ at different strains can be estimated by alternative approximated finite displacement difference method rather than IFCs calculations~\cite{phonopy}. For strain-free monolayer InSe, it gives $\gamma$=0.53 which agrees with $\gamma$=0.50 obtained by calculation of $2^nd$ and $3^rd$ IFCs. The frequency of phonon and force constants are known to have a negative and positive correlation with $\gamma$~\cite{expansion}, leading to complex behavior of $\gamma$ with respect to strain $\varepsilon$. In monolayer InSe, we found that $\gamma$ increases as tensile $\varepsilon$ increase (see Fig.~\ref{wh3}(b)). Thus, the decrease of $\theta_{D}$ and increase of $\gamma$ lead to the decrease of $\kappa$ of monolayer InSe under tensile strain according to Slack mode~\cite{slack1973nonmetallic}. That is similar to many bulk materials~\cite{bulkstrain} and other 2D materials such as MoS$_{2}$~\cite{MoS2strain2}. On the other side, the compressive strain will decrease the $\kappa$ of 2D InSe.
We have also calculated the $\kappa$ of monolayer InSe under tensile strain of $\epsilon$=2.0\% and $\epsilon$=4.0\% through more exact phonon BTE equation (see Fig.~\ref{wh3}(b)). It was found that $\kappa$ was reduced by 11\% and 18\% with respect to $\kappa$ at strain-free case, which is consistent with the estimation by Slack model.

\subsubsection{Size effects}
Considering the limited sample size (L), the size effects is another method to modulate $\kappa$. Here we consider the boundary scattering in the range of diffusive thermal transport, which corresponds $1/\tau_{\mathbf{q},s}=1/ \tau_{\mathbf{q},s}^{an}+1/\tau_{\mathbf{q},s}^{iso}+1/\tau_{\mathbf{q},s}^{b}$. The
result of monolayer group-III chalcogenides are displayed in Fig. S5. At room temperature, the phonon mean free
path (MFP) of monolayer group-III chalcogenides is about $10^{5}$ nm, which is mainly determined by the MFP of ZA and LA phonons. The $\kappa$ can be decreased by 90\% as the L decrease down to 10 nm (see Fig. S5), indicating that nanostructuring might be an effective method to reduce the $¦Ê\kappa$ of 2D group-III chalcogenide, consistent to previous work~\cite{InSe}.

\begin{figure}[tbp!]
\centerline{\includegraphics[width=0.6\textwidth]{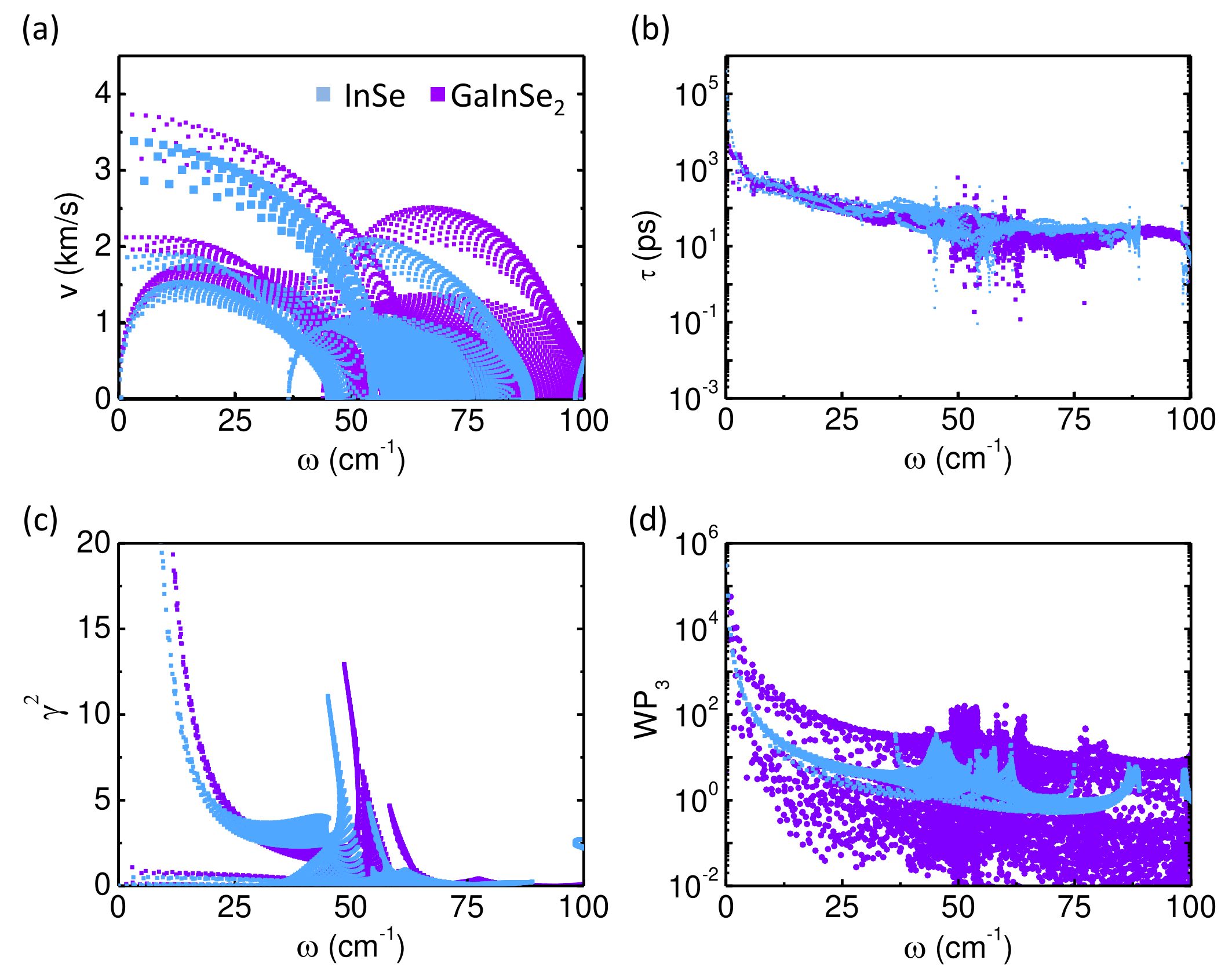}}
\caption{The frequency dependence of (a) mode phonon group velocity $v_{\mathbf{q},s}$, (b) the phonon lifetime $\tau_{\mathbf{q},s}$, (c) the square of  gr\"{u}neisen parameter $\gamma_{\mathbf{q},s}^{2}$ and (d) weighted phase space $W_{\mathbf{q},s}$ for monolayer InSe and GaInSe$_{2}$.}
\label{wh4}
\end{figure}

\subsubsection{Janus structures}
Recently, a new Janus-type structure, monolayer MoSSe, has been synthesized through the
replacement of the S atoms at one side of monolayer MoS$_{2}$ by Se atoms or Se atoms of monolayer MoSe$_{2}$ by S atoms~\cite{janus1}. The $\kappa$ of monolayer MoSSe is between that of monolayer MoS$_{2}$ and MoSe$_{2}$~\cite{C8CP00350E}.
Motivated by that, the effort has been denoted into the electronic~\cite{InSeS}, piezoelectric~\cite{piezoelectric} and valleytronic~\cite{Valley} properties of Janus structures based on 2D group-III chalcogenide.
In a previous work, we have shown that monolayer In$_{2}$SSe has a higher $\kappa$ but a lower $\mu$ than monolayer InSe~\cite{Wan_2019}. On the other side, monolayer In$_{2}$SeTe has a higher $\mu$ but a lower $\kappa$ than InSe. Here, we considered anther 2D Janus structure InGaSe$_{2}$.
The cohesive energy of monolayer InGaSe$_{2}$ is -3.43 eV/atom, larger than InSe (-3.34 eV/atom) and smaller than GaSe (-3.58 eV/atom). Meanwhile, Indium atoms of monolayer InSe are also on the surface, so Janus InGaSe$_{2}$ may could be formed by replacing the In atoms of one side of InSe by Ga atoms. The phonon dispersion ensures its structural stability (see Fig. S1(e)).

Compared to the continuously modulation of $\kappa$ by tensile strain and rough boundary scattering, the change of ¦Ê due to Janus structures can keep permanent but  dependents on the structures details. The phonon group velocity $v_{\mathbf{q},s}$ of monolayer GaInSe$_{2}$ is larger than InSe, due to smaller atomic mass (see Fig.~\ref{wh4}(a)). GaInSe$_{2}$ occurs a breaking of mirror symmetry compared to InSe. This will lead to asymmetric chemical bonding and charge density distribution along the z-direction, which will increase bond anharmonicity~\cite{C7CP02486J,bonding}.
Fig.~\ref{wh4}(c,d) show the square of mode Gr\"{u}neisen parameter $\gamma_{\mathbf{q},s}^{2}$ and weighted phase space $W_{\mathbf{q},s}$, which represents the anharmonicity strength and the number of channels available for phonon scattering, respectively~\cite{weightspace,li2014shengbte}. Both $\gamma_{\mathbf{q},s}^{2}$ and $W_{\mathbf{q},s}$ of monolayer GaInSe$_{2}$ is larger than InSe, lead to stronger phonon scattering and smaller $\tau_{\mathbf{q},s}$ of phonon modes than that of monolayer InSe at energy zone $\omega>$25 cm$^{-1}$.
As a result, GaInSe$_{2}$ has overall higher $v_{\mathbf{q},s}$ and lower $\tau_{\mathbf{q},s}$ than InSe. The competition between $v_{\mathbf{q},s}$ and $\tau_{\mathbf{q},s}$ leads to that monolayer GaInSe$_{2}$ has a $\kappa$ comparable to that of monolayer InSe (see Fig.~\ref{wh1}(d)).

\begin{figure}[tbp!]
\centerline{\includegraphics[width=0.6\textwidth]{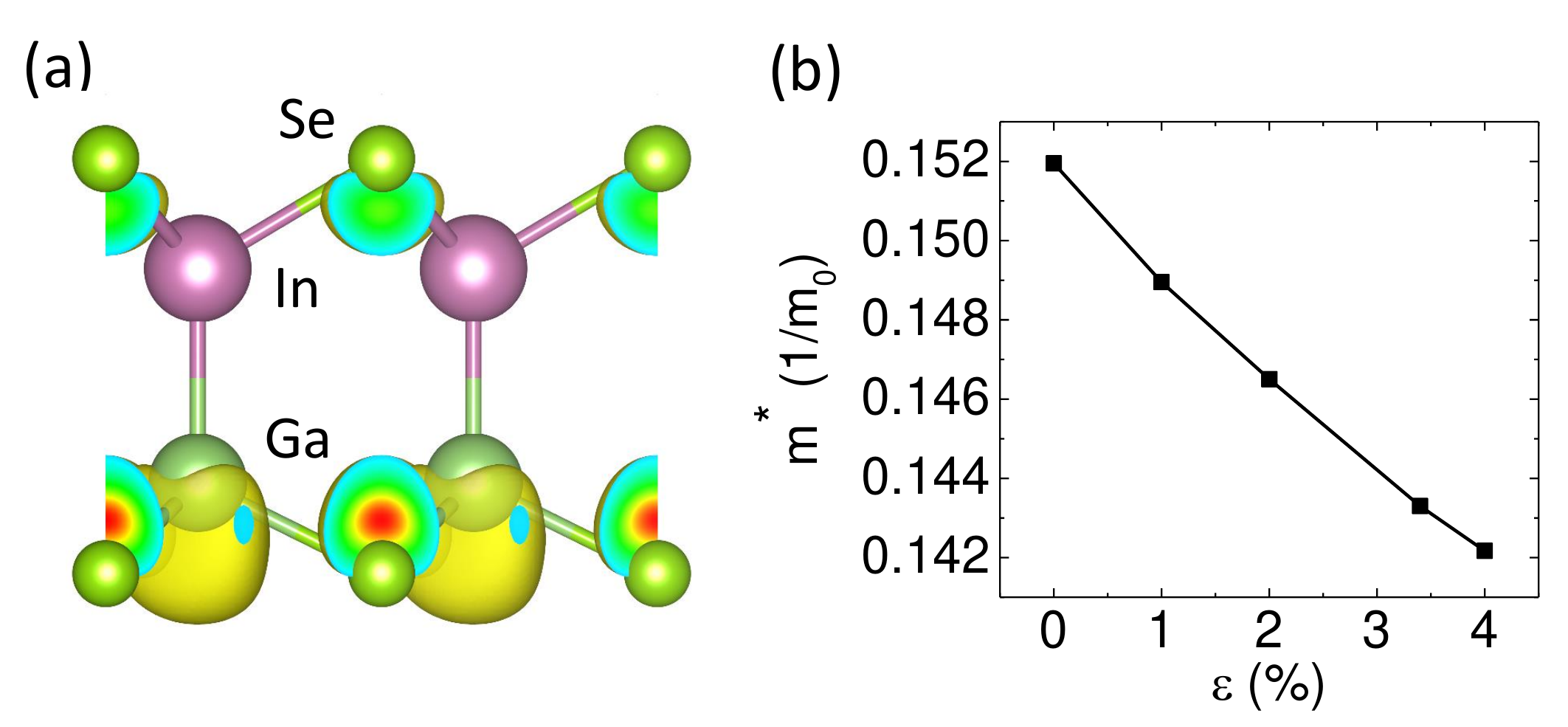}}
\caption{(a)The spatial distribution of electronic state at the conduction band minimum of monolayer GaInSe$_{2}$. (b) The electron effective mass ($m^{*}$) as a function of strain.}
\label{wh5}
\end{figure}

\Table{\label{table2}The electron effective mass $m^{*}(1/m_{0})$ along $x$ and $y$ axis, 2D elastic module $C_{\rm 2D}$ (J/m$^{2}$), deformation potential constant $E_{1}$ (eV) and room-temperature $\mu$ (cm$^{2}$/V/s) of monolayer InSe and GaInSe$_{2}$.}
\br
Type    &$m^{*}_{x}/m_{0}$ &$m^{*}_{y}/m_{0}$ &$C_{\rm 2D_x}$ &$|E_{1x}|$ & $\mu_{x}$\\
\mr
InSe            &0.181  &0.182    &49.21          &5.815   &943.3 \\
GaInSe$_{2}$    &0.160	 &0.160   &56.76          &6.537   &1107.4 \\
\br
\endTable

On the other side, the electron mobility $\mu$ of monolayer GaInSe$_{2}$ was calculated based on the deformation potential theory~\cite{deformation}. The calculation procedure has been explained in previous work~\cite{Wan_2019}. The calculated parameters involved in $\mu$ are shown in Table~\ref{table2}. The 2D effectively elastic modulus $C_{2D}$ of monolayer GaInSe$_{2}$ is larger than that of monolayer InSe. That is consistent with that shorter bonding length $d_{ac}$ and $d_{cc}$ of monolayer GaInSe$_{2}$ than monolayer InSe (see Table. S1). The electron effective mass $m^{*}$ of monolayer GaInSe$_{2}$ is smaller than that of monolayer InSe. The orbital analysis indicates that In-$5s$ orbitals dominate the electronic states at conduction band minimum (CBM) of monolayer InSe. In contrast, the CBM of monolayer GaInSe$_{2}$ is mainly composed of 5s
orbital of Ga atom as well as Se-$p_{z}$ orbital (see Fig.~\ref{wh5}(a)), due to electric potential difference arising from the charge transfering from In to Ga atoms. The Ga side of monolayer GaInSe$_{2}$ experiences a tensile strain about 3.4\% compared to that of monolayer GaSe, seen from the lattice constant in Table. S1. The $m^{*}$ of GaSe will decrease as the increase of tensile strain (see Fig.~\ref{wh5}(b)). At a tensile strain of 3.4\%, the $m^{*}$ is 0.144 $m_{0}$, consistent with small $m^{*}$ of GaInSe$_{2}$ (see Table.~\ref{table2}). Thus, monolayer GaInSe$_{2}$ has a smaller $m^{*}$ and larger $C_{2D}$ and deformation potential $E_{l}$, leading to a higher $\mu$ than that of InSe. Meanwhile, considering its $\kappa$ is comparable to monolayer InSe, Janus monolayer GaInSe$_{2}$ is superior to monolayer InSe, monolayer In$_{2}$SSe and In$_{2}$SeTe in the electronic applications.

\section{Conclusion}
Based on the first-principles calculations, we investigated the modulation of lattice thermal conductivity of monolayer group-III chalcogenides. The room-temperature thermal conductivity of monolayer GaS, InS, GaSe and InSe is 111.0, 83.10, 58.9 and 44.9 W/(mK), respectively, which is desirable for heat dissipation of relevant devices. The heat transport of group-III chalcogenides can be well described by Slack mode and are dominated by phonon group velocity. Both acoustic branches and low-frequency optical branches have contribution to thermal transport of group-III chalcogenides. Based on that, we proposed three feasible methods to modulate the $\kappa$. We found that the $\kappa$ of 2D group-III chalcogenides could be continuously decreased by tensile strain and rough boundary scattering but be increased by compressive strain. The change of $\kappa$ due to Janus structures is permanent which depends on the structures details. Taking 2D InSe as an example, among the various Janus structure we found that the Janus monolayer GaInSe$_{2}$ could keep a comparable $\kappa$ and a higher electron mobility than monolayer InSe.
Our work helps to understand of the thermal transport of 2D group-III chalcogenides and provides a guide for the thermal management in the devices based on these materials.

\section{Acknowledgments}
This work was supported by National Natural Science Foundation of China (No.11904312 and 11904313), the Project of Hebei Educational Department, China (No.ZD2018015 and QN2018012), the Advanced Postdoctoral Programs of Hebei Province (No.B2017003004) and the Natural Science Foundation of Hebei Province (No. A2019203507).
The numerical calculations in this paper have been done on the supercomputing system in the High Performance Computing Center of Yanshan University.

\clearpage

\bibliographystyle{unsrt}
\bibliography{thinse1}

\end{document}